\documentclass[aps
,prl,10pt
  ,twocolumn,
twoside,
showpacs,
 nobalancelastpage,
superscriptaddress
,floatfix
 ]{revtex4}

\usepackage{bm}
\usepackage{amsfonts}
\usepackage{amsmath}
\usepackage{amssymb}
\usepackage[latin2]{inputenc}
\usepackage[T1]{fontenc}
\usepackage{epsfig}
\usepackage{mathrsfs}
\usepackage{verbatim}
\usepackage{latexsym}

\newcommand{\proof}{{\it Proof: }}
\newcommand{\proofend}{$\Box$}

\newcommand{\il}[2]{\big<#1\big|#2\big>}

\newcommand{\ket}[1]{|#1\rangle}
\newcommand{\bra}[1]{\langle#1|}

\newcommand{\braket}[2]{\big<#1\big|#2\big>}

\newcommand{\be}[0]{\begin{equation}}
\newcommand{\ee}[0]{\end{equation}}
\newcommand{\bea}[0]{\begin{eqnarray}}
\newcommand{\eea}[0]{\end{eqnarray}}

\newcommand{\e}{\text{e}}
\newcommand{\tr}{\text{tr}}

\newtheorem{thm}{Theorem}
\newtheorem{df}{Definition}
\newtheorem{lem}{Proposition}

\newlength{\dinwidth}
\newlength{\dinmargin}
\DeclareMathAlphabet{\scr}{U}{rsfs}{m}{n}

\begin{document}
\title{Subadditivity of the minimum output entropy and superactivation of the classical capacity of quantum multiple access channels}

\author{Ł. Czekaj}
\affiliation{Faculty of Applied Physics and Mathematics,
Gda{\'n}sk University of Technology, 80-952 Gda{\'n}sk, Poland}
\affiliation{National Quantum Information Center of Gda\'nsk, 81-824 Sopot, Poland\\
lczekaj@mif.pg.gda.pl}

\begin{abstract}
We study subadditivity of the minimum output entropy ($H_{min}$) of quantum multiple access channels (MACs). We provide an example of violation of the additivity theorem for $H_{min}$ known in classical information theory. Our result is based on a fundamental property of $MACs$ i.e. independence of each sender. The channels used in the example can be constructed explicitly. On the basis of subadditivity of $H_{min}$ we also provide an example of extremal superadditivity (super activation) of the classical capacity region of MACs.
\end{abstract}


\maketitle

Using quantum resources like quantum entanglement \cite{quant_ent_big} in quantum information theory \cite{chuang_nielsen} leads to a new class of effects, known as {\it quantum activation}, which are impossible in classical information theory \cite{coverthomas}.
Some examples of quantum activation are: (i) superadditivity of the classical capacity $\mathcal{C}$ in the fundamental case of 1-to-1 quantum channels \cite{hastings} where transmission of entangled states leads to capacities larger than using product states, (ii) nonlocality effect for classical capacity region $\mathcal{R}$ of quantum multiple access channels \cite{disc_supp,cont_supp} where entanglement used by first sender increases maximal rate of an other ($R_2$) without increasing the maximal total rate $R_1+R_2$. The effect can be quite strong as shown in \cite{grudka}, (iii) superactivation, i.e. extremal superadditivity, of the quantum capacity $\mathcal{Q}$ for 1-to-1 channels where two quantum channels with $0$ quantum capacities working together allow for transmission of qubits \cite{yard_smith}.

This paper addresses subadditivity of the minimum output entropy $H_{min}$ and quantum activation of the classical capacity region $\mathcal{R}$ of the entanglement breaking \cite{ebc} multiple access channels (MACs) and it continues the research started in Ref.~\cite{disc_supp}.
Effect of the subadditivity of $H_{min}$ is manifested when transmission of entangled states produces lower entropy than transmission of any product states.
The question of subadditivity of $H_{min}$ of quantum 1-to-1 channels appears to
have been first considered in print in Ref.~\cite{min_entropy}. In Ret.~\cite{additivity_equiv} the equivalence between additivity of $H_{min}$ and classical capacity $\chi$ was stated. Finally, an example of subadditivity of $H_{min}$ for quantum 1-to-1 channels was first provided in Ref.~\cite{hastings}.
 and explored further in \cite{horodecki_notes_hastings}. Hastings' channels seems to be very hard to explicitly construct since this task requires a search through the set of unitary matrices.
Here we study the subadditivity of $H_{min}$ in the realm of MACs. The setup we present is intrinsically MAC and cannot be reduced to the setup of 1-to-1 channels, i.e. to the case studied by Hastings.
The advantage of our approach is the existence of effective algorithms  allowing the explicit construction of the channels we present.
It should allow a better understanding of the mechanism behind the subadditivity effect.
The subadditivity of $H_{min}$ leads us to the quantum activation of $\mathcal{R}$.
The example provided here exhibits superadditivity of the total rate $R_T$. We construct two sequences of channels $\{\tilde{\Gamma}_A^{(\delta)}\},\{\tilde{\Gamma}_B^{(\delta)}\}$ and study its parallel setup $\{\tilde{\Gamma}_A^{(\delta)}\}\otimes\{\tilde{\Gamma}_B^{(\delta)}\}$. Without using entanglement in  communication, $R_T\rightarrow 0$ as $\delta\rightarrow 0$. On the other hand, using entangled states allows to achieve $R_T=1$ for each $\delta$. This can be view as a superactivation effect, since entanglement strongly activates channels with almost $0$ capacities.

The superadditivity of $\mathcal{C}$ in entanglement breaking MACs suggests qualitative differences between bipartite and multipartite communication since it cannot occur for entanglement breaking 1-to-1 channels. It was first pointed out in Ref.~\cite{grudka}. Superadditivity was shown for the entanglement breaking MACs cooperating with an identity channel (which is not entanglement breaking). Here we move one step further and show that the very strong superadditivity takes place also if we use only entanglement breaking MACs.

The paper is organized as follows: first we provide definitions and theorems used in the main part of the paper. We stress on the explanation of the idea of a randomness extractor which is of paramount importance to further considerations. Then we present the main results, i.e. subadditivity of $H_{min}$ and superactivation of the classical capacity of the MACs.

\section{background}
A {\it quantum channel} $\Gamma$ is a linear, completely positive and trace preserving map from density matrices to density matrices $\rho \mapsto \Gamma(\rho)$ \cite{chuang_nielsen} and it models transmission of quantum states in the presence of noise. An {\it entanglement breaking channel} is a quantum channel which cannot be used to create entanglement between parts participating in communication \cite{ebc}. It can be presented in the form of a measurement followed by a state preparation. In quantum {\it multiple access channels} there are at least two senders transmitting to one receiver. Each sender sends his state independently of the other, i.e. their inputs are uncorrelated. For the case of two senders, a MAC acts as a map:
\begin{equation}
\rho_1\otimes\rho_2\mapsto\Gamma(\rho_1\otimes\rho_2)
\end{equation}
where state $\rho_1$ ($\rho_2$) is sent by the sender $S_1$ ($S_2$).

We will denote as $\Gamma_A\otimes\Gamma_B$ a {\it parallel setup} of channels $\Gamma_A,\!\Gamma_B$. It means that each sender
has access to one input of the channel $\Gamma_A$ and one input of the channel $\Gamma_B$. They can transmit any states through their inputs where the first part of transmitted states goes through $\Gamma_A$ and the second through $\Gamma_B$.
Channels are used synchronously and the receiver has access to the outputs of both channels.

A quantum channel can be used for transmission of either classical \cite{class_cap} or quantum information \cite{quant_cap}.
In the {\it transmission of classical information}, senders encode classical messages $\{i\},\{j\}$ into {\it code states} transmitted through the channel $i \mapsto\rho_1^{(i)},\! j \mapsto \rho_2^{(j)}$. Senders and receiver know the ensemble of code states (i.e. the set of code states and the probabilities the states are transmitted with) $\{p_1^{(i)},\rho_1^{(i)}\}$, $\{p_2^{(j)},\rho_2^{(j)}\}$ but one sender does not know which state is transmitted by the other sender at a given time. The receiver performs measurement on the output state and based on its result tries to infers which message $(i,j)$ was transmitted.

The amount of classical information which can be reliablly transmitted through the MAC is given by the pair of rates $(R_1,R_2)$. The set of all achievable rates form the Holevo like {\it classical capacity region} $\mathcal{R}(\Gamma)$.
For a given ensemble of code states one can define the state $\rho=\sum_{i,j}p_1^{(i)}p_2^{(j)}e_1^{(i)}\otimes e_2^{(j)}\otimes\Gamma(\rho_1^{(i)}\otimes\rho_2^{(j)})$ where $\{e_1^{(i)}\}$,($\{e_2^{(j)}\}$) are projectors on the standard basis of the Hilbert space of the input controlled by $S_1$ ($S_2$).
%
The capacity region $\mathcal{R}(\Gamma)$ is obtained as a convex closure of all rates $(R_1,R_2)$ such that there exists $\rho$ for which the set of inequalities is fulfilled:
\begin{eqnarray}
&R_1 \leq I(S_1:R|S_2)&\\
&R_2 \leq I(S_2:R|S_1)&\\
&R_T=R_1+R_2 \leq I(S_1,S_2:R)&
\end{eqnarray}
where $I(S_1,S_2:R) = H(\rho_{S_1})+H(\rho_{S_2})-H(\rho_{R})$ and $I(S_1:R|S_1)=\sum_j p_j I(S_1:R|S_2=j)$. $H(\rho)= -\tr [\rho \log\rho]$ is the von Neuman entropy. $R_T$ denotes the total rate and is defined as $R_T=\sum_i R_i$.
We shall denote "single shot" formula $\mathcal{R}^{(1)}(\Gamma)=\mathcal{R}(\Gamma)$ and $\mathcal{R}^{(n)}(\Gamma)=\frac{1}{n}\mathcal{R}(\Gamma^{\otimes n})$ for the situation where code states can be $n$-particle entangled states. The interesting case is that of the regularized capacity region $\mathcal{R}^{(\infty)}(\Gamma)=\lim_{n \rightarrow \infty} \frac{1}{n}\mathcal{R}^(\Gamma^{\otimes n})$, which express the upper bound for the capacity region which can be achieved due to quantum entanglement.

For a MAC $\Gamma$ with $n$ senders we define the {\it minimum output entropy} $H_{min}(\Gamma)$ as:
\begin{equation}
H_{min}(\Gamma)=\min_{\rho_{1},\ldots,\rho_n} H(\Gamma(\rho_{1}\otimes\ldots\otimes\rho_{n}))
\end{equation}
where $\rho_i$ belongs to the input space of the sender $S_i$.  Minimization runs over all states from the input space of each sender. Due to the concavity of $H(\rho)$, it is sufficient to minimize only over pure states.

In the classical setup senders transmit only product states from orthogonal bases of the input spaces of $\Gamma_A$ and $\Gamma_B$.
By the properties of the von Neuman entropy, we can state for MACs the additivity theorem:
\begin{equation}
H_{min}(\Gamma_A)+H_{min}(\Gamma_B) = H_{min}(\Gamma_A\otimes\Gamma_B)
\end{equation}
Existence of entangled states in the input space of the $\Gamma_A\otimes\Gamma_B$ extends the set we minimize $H_{min}$ over and makes the additivity theorem invalid in the quantum setup. {\it Subadditivity of $H_{min}$} occurs if transmission of entanglement state through $\Gamma_A\otimes\Gamma_B$ produce lower entropy than the sum of $H_{min}$ of each channel working separately.

The additivity theorem for MACs can be stated analogically:
\begin{equation}
\mathcal{R}(\Gamma_A)+\mathcal{R}(\Gamma_B) = \mathcal{R}(\Gamma_A\otimes\Gamma_B)
\end{equation}
Here we use the geometrical sum of the sets in Euklides space.
In the parallel setup of quantum MACs we can use entangled states as code words.
 Superadditivity of the classical capacity regions $\mathcal{R}$ takes place if there exists a protocol using entangled code states with classical capacity region $\mathcal{R}_{ent}(\Gamma_A\otimes\Gamma_B)$ such that for each protocol using only product code states with  $\mathcal{R}_{prod}(\Gamma_A\otimes\Gamma_B)=\mathcal{R}(\Gamma_A)+\mathcal{R}(\Gamma_B)$ occurs $\mathcal{R}_{prod}\subsetneq\mathcal{R}_{ent}$.
Superactivation describes the situation when $\mathcal{R}_{ent}$ is huge in comparison with $\mathcal{R}_{prod}$.

In what follows, we shall use generalized Bell states \cite{gen_bell_states} in the form:
\begin{equation}
\ket{\psi_{ \alpha,\beta}} = \frac{1}{\sqrt{D}}\sum_{l=0}^{D-1} \exp \left( \frac{2 \pi i}{D} \alpha l \right) \ket{l}\ket{l+\beta}
\end{equation}
$\alpha,\beta \in\{0,D-1\}$ are indices. The states belong to the space  $\mathbb{C}^D\otimes\mathbb{C}^D$.
For the state $\ket{\mu}=\sum_i\mu_i\ket{i}$ where $\{\ket{i}\}$ is the standard basis, we will write $\ket{\mu^*}=\sum_i\mu^*_i\ket{i}$ and $D=2^d$.

For two random variables $X,Y$ with equal support the statistical distance is defined as:
\begin{equation}
\mathrm{dist}(X,Y)=\frac{1}{2}\sum_{e\in \mathrm{sup}(X)}|p_x(e)-p_y(e)|
\end{equation}
We denote by $F_m$ a $m$-bit random variable with a flat distribution over its support.

\begin{lem}
For a binary random variable $X$ if $\mathrm{dist}(X,F_2)=\epsilon$ then $1-H(X)=(2/\ln 2)\epsilon^2 + O(\epsilon ^4)$ and for $\epsilon \in (0,0.5)$ we have $1-H(X)\leq 4\epsilon^2$.
\label{dist_entropy_lem}
\end{lem}
\proof This follows directly from a Taylor series expansion.

A classical {\it multiple source randomness extractor} is a function which distills entropy from independent "week random sources" into random variables with almost flat distribution. We put the word "classical" to distinguish from situation where randomness is obtained on the base of quantum effects.

Usability of a random source in a randomness extraction process is characterized by the {\it min-entropy} $H_\infty$ \cite{Chor_goldreich, Zuckerman} defined as:
\begin{equation}
H_{\infty}(X)=\min_{x\in\text{sup}X} - \log p(x)
\end{equation}

\begin{df}{Multiple source randomness extractor \cite{extractrors_bounds}}:
A function $f_\mathrm{ext}: \{0,1\}^{n\times l} \mapsto \{0,1\}^m$ which satisfies:
\begin{equation}
\mathrm{dist}\left( f_\mathrm{ext} \left( X_1,\ldots,X_l \right), F_m \right)\leq \epsilon
\end{equation}
for every independent $n$-bit source $X_1,\ldots,X_l$ with $H^\infty(X_i)\geq k$
is called an $l$-source extractor with $k$ min-entropy requirement, $n$-bit input, $m$-bit output and $\epsilon$-statistical distance.
\end{df}

\begin{thm}{Extractor existence \cite{extractrors_bounds}}:
Let $m<k<n$ be integers and let $\epsilon>0$.
If $k > \log n + 2 m + 2 \log ( 1/\epsilon ) + 1 $ holds then there exists a 2-source extractor $f_\mathrm{opt}: \{0,1\} ^{n\times 2}\rightarrow \{0,1\}^m$ with k-entropy requirement and distance $\epsilon$.
The extractor can be computed in time proportional to $2^{5n^2 2^{2k}}$.
\label{extr_exist_lem}
\end{thm}

\section{Subadditivity of $H_{min}$}
Here we provide two families of MACs $\{\Gamma_A^{(\delta)}\}$ and $\{\Gamma_B^{(\delta)}\}$ indexed by $\delta$ which exhibit subadditivity of $H_{min}$ for $\delta<1/2$.

\begin{figure}
	\centering
		\includegraphics[scale=0.45]{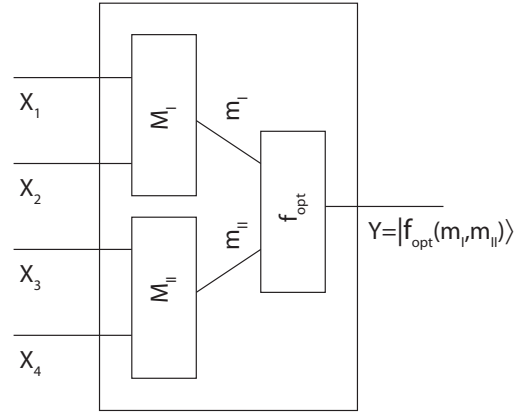}
	\caption{The general schema of the channels $\Gamma_A$ and $\Gamma_B$. $X_i$ are $d$-qubits input lines. $M_{I}$ and $M_{II}$ are measurements. Its result is denoted by $m_{I}$ and $m_{II}$ respectively. $f_\mathrm{opt}$ is a classical randomness extractor with properties described by the Thm.~\ref{extr_exist_lem}. }
	\label{fig:gamma}
\end{figure}

The channels $\Gamma_A^{(\delta)},\Gamma_B^{(\delta)}$ consist of $4$ independent $d$-qubit inputs $X_1^{A},\ldots,X_4^{A}$ ($X_1^{B},\ldots,X_4^{B}$) and $1$-qubit output $Y^{A}$ ($Y^B$). The inputs $X_i^{A}$ and $X_i^{B}$ are controlled by the sender $S_i$. The size of the inputs depend on $\delta$ as follows: $d = \left\lceil 2\log (1/\delta)+12\right\rceil$.
Both channels are based on the same scheme (see FIG. \ref{fig:gamma}) so we will describe the channel $\Gamma_A$ and point out where the channels differ. In the first step, the channel perform measurements $M_I$ and $M_{II}$. $M_I$ is a joint measurement on inputs $X_1,X_2$ and $M_{II}$ is a joint measurement on inputs $X_3,X_4$. In the channel $\Gamma_A$, measurements $M_I$ and $M_{II}$ are performed in the basis $\{\ket{\Psi_{\alpha,\beta}}\}$ while in the channel $\Gamma_B$ in the basis $\{\ket{\Psi^*_{\alpha,\beta}}\}$. The result of the measurement $M_{I}$ ($M_{II}$) is denoted by $m_I$ ($m_{II}$). $m_I$ and $m_{II}$ provide the $2d$-bits inputs to the randomness extractor $f_\mathrm{opt}$, which produces a $1$ bit output. Existence of the extractor $f_\mathrm{opt}$ with proper features will be proven later. Depending on the value of $f_\mathrm{opt}(m_{I},m_{II})$, the channel produces the output state $\ket{0}$ or $\ket{1}$.

\begin{figure}
	\centering
		\includegraphics[scale=0.45]{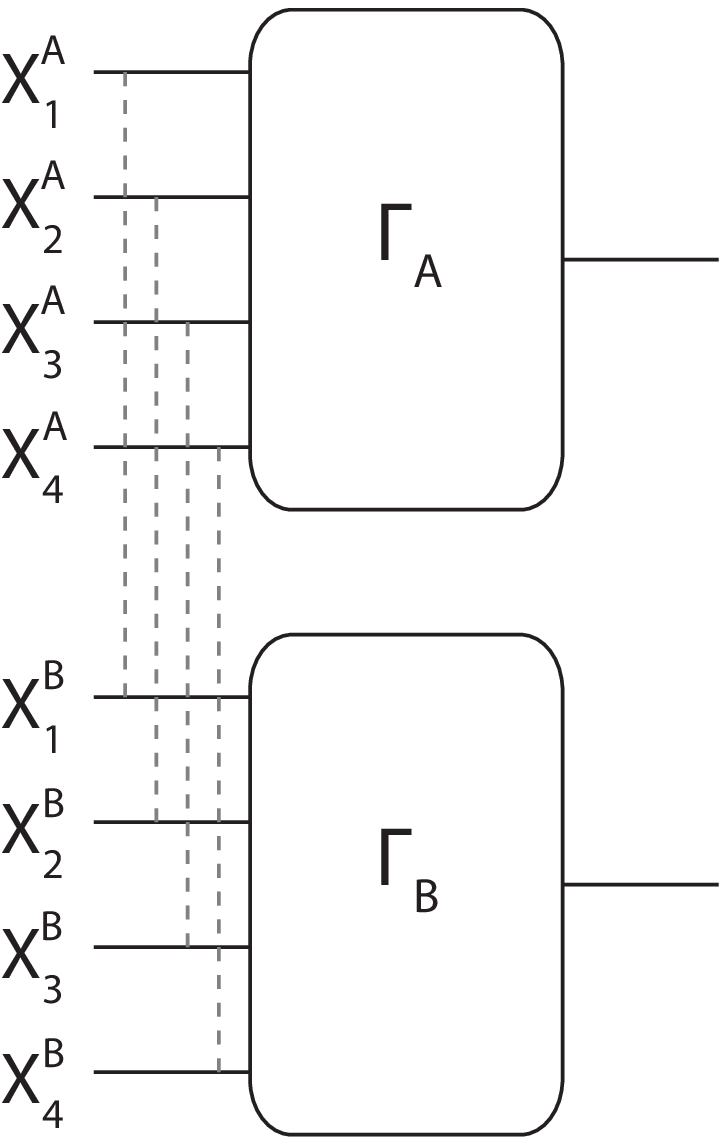}
	\caption{The parallel setup of the channels $\Gamma_A$ and $\Gamma_B$. Dashed lines depict entanglement of the inputs in case of $\ket{\Psi_{0,0}}$ transmission through $\Gamma_A\otimes\Gamma_B$. }
	\label{fig:parallel}
\end{figure}

We will show that for any $\delta>0$, $H_{min}(\Gamma_A^{(\delta)})$ and $H_{min}(\Gamma_B^{(\delta)})$ cannot be lower than $1-\delta$. On the other hand we will show that if each sender transmits $\ket{\Psi_{0,0}}$, the output entropy of the $\Gamma_A^{(\delta)}\otimes\Gamma_B^{(\delta)}$ is equal to $1$. Since this is the upper bound for $H_{min}(\Gamma_A^{(\delta)}\otimes\Gamma_B^{(\delta)})$ we will prove that for $\delta<1/2$,  $H_{min}(\Gamma_A^{(\delta)})+H_{min}(\Gamma_B^{(\delta)})>H_{min}(\Gamma_A^{(\delta)}\otimes\Gamma_B^{(\delta)})$.

We start by proving that $H_{min}(\Gamma_A^{(\delta)})=H_{min}(\Gamma_B^{(\delta)}) \geq 1-\delta$ but first we give
a proposition which will be useful in what follows.

\begin{lem}
For the random variables associated with the outputs of the measurements performed by the channels $\Gamma_A^{(\delta)}$ and $\Gamma_B^{(\delta)}$, the following holds:
$H^\infty(M_I^A)=H^\infty(M_{II}^A)=H^\infty(M_I^B)=H^\infty(M_{II}^B)=d$. Here $d$ denotes the input size of the channels.
\label{H_inft_pre}
\end{lem}

\proof
We will only prove that $H^\infty(M_I^A)=d$ since the other cases can be proved analogously.
Let the projector measurement $M_I^A$ be
performed on the product state $\ket{\mu}\otimes\ket{\nu}$, where  $\ket{\mu}=\sum_{j=0}^{D-1} \mu_j \ket{j},\:\ket{\nu}=\sum_{k=0}^{D-1} \nu_k \ket{k}$ are $d$-qubit states pertaining to senders $S_1$ and $S_2$ respectively.
We will show that the probability $p(\alpha,\beta)=|\il{\psi_{m,n}}{\mu}\ket{\nu}|^2$
of getting the pair $(\alpha,\beta)$ as the result of measurement satisfies $p(\alpha,\beta)\leq \frac{1}{D}$. $H_{\infty}(M_I^A)\geq d$ is simple consequence of this fact.

Observe that:
\begin{eqnarray}
p(\alpha,\beta)&=&\il{\psi_{\alpha,\beta}}{\mu}\ket{\nu}\\
&=&\frac{1}{\sqrt{D}} \sum_{l=0}^{D-1} \exp \left( \frac{2 \pi i}{D} \alpha l \right) \bra{l}\bra{l+\beta} \\
&&\sum_{j=0}^{D-1} \mu_j \ket{j}\sum_{k=0}^{D-1} \nu_k \ket{k}\nonumber\\
&=&\frac{1}{\sqrt{D}} \sum_{j=0,k=0,l=0}^{D-1}
\exp \left( \frac{2 \pi i}{D} \alpha l \right) \\
&&\mu_j \nu_k\braket{l}{j}\braket{l+\beta}{k}\\
&=&\frac{1}{\sqrt{D}}\sum_{l=0}^{D-1}\exp \left( \frac{2 \pi i}{D} \alpha l \right) \mu_l \nu_{l+\beta}\\
&=&\frac{1}{\sqrt{D}}\bra{\mu^*}U_\beta^\alpha\ket{\nu} \label{scalarprodres}\\
\end{eqnarray}
where $U_{\beta}^{\alpha}=\sum_{l=0}^{D-1}\ket{l+\beta}\bra{l}\exp \left( \frac{2 \pi i}{D} \alpha l \right)$ is an unitary.
Finally by the property of the scalar product we have $p(\alpha,\beta)=\frac{1}{D}|\bra{\mu^*}U_\beta^\alpha\ket{\nu}|^2\leq 1$.
\proofend

Taking into account proposition \ref{H_inft_pre},
and noting that $d/2>\log d$ is true for $d > 4$, we obtain that $d = \left\lceil 2\log 1/\delta+12\right\rceil$ fulfills the
requirements of Thm.~\ref{extr_exist_lem} with $\epsilon = \sqrt{\delta}/2$. Since $f_\mathrm{opt}$ exists and has statistical distance $\epsilon$, by proposition \ref{dist_entropy_lem} for each channel we have $H_{min}\geq 1-\epsilon$.

Now consider output entropy of the $\Gamma_A^{(\delta)}\otimes\Gamma_B^{(\delta)}$ if all senders transmit $\ket{\psi_{0,0}}$. The first part of the $2d$-qubit state is transmitted through the channel $\Gamma_A^{(\delta)}$ and the second through $\Gamma_B^{(\delta)}$ (see FIG. \ref{fig:parallel}).
We will show that in this case the output entropy of the $\Gamma_A^{(\delta)}\otimes\Gamma_B^{(\delta)}$ cannot exceed $1$.

The randomness extractor $f_\mathrm{opt}$ is a deterministic function of the outcome of the measurements $M_I, M_{II}$. Its output controls which of the pure states $\ket{0},\ket{1}$ will be the output of the channel. If the results of measurements $M_{I}^A$ and $M_{I}^B$ ($M_{I}^A$ and $M_{I}^B$) are identical, then so will the outputs of the channels too. Let us focus on the measurements $M_{I}^A$ and $M_{I}^B$. We will show that $p(m^A_I,m^B_I)=p(\alpha_A,\beta_A,\alpha_B,\beta_B)\propto\delta_{\alpha_A,\alpha_B}\delta_{\beta_A,\beta_B}$.

\begin{eqnarray}
&&p(\alpha_A,\beta_A,\alpha_B,\beta_B)\\
&=&\frac{1}{D^2} \left|\bra{\psi_{0,0}^*} U_{\beta_A}^{\alpha_A}\otimes U_{\beta_B}^{\alpha_B \dagger}\ket{\psi_{0,0}}\right|^2\label{measurment_trick}\\
&=&\frac{1}{D^2}\left|\frac{1}{D} \sum_{k,l=0}^{D-1}\bra{k}\bra{k} \exp\left[ i\frac{2 \pi}{D} (\alpha_A - \alpha_B) l \right]\right.\\
&&\left. \ket{l+\beta_A}\ket{l+\beta_B} \right|^2 \\
&=&\frac{1}{D^4}\left| \sum_{l}^{D-1} \exp\left[ i\frac{2 \pi}{D} (\alpha_A - \alpha_B) l \right] \delta_{\beta_A,\beta_B} \right|^2\\
&=&\frac{1}{D^2}\delta_{\alpha_A,\alpha_B}\delta_{\beta_A,\beta_B}
\end{eqnarray}
where Eq.~(\ref{measurment_trick}) is obtained in the same way as Eq.~(\ref{scalarprodres}). This result can be viewed as generalized entanglement swapping \cite{gen_bell_states,swap_1,swap_2}. Entanglement between uses of channels $\Gamma_A$ and $\Gamma_B$ is swapped by the measurement $M_{I}^A$ into entanglement between the inputs of the channel $\Gamma_B$ belonging to senders $S_1$ and $S_2$. The argument presented is also valid for the measurements $M_{II}^A$ and $M_{II}^B$.

As we have shown, in the case of entangled state transmission, the outputs of $f_\mathrm{ext}$ for channels $\Gamma_A$ and $\Gamma_B$ are equal. Since the output of the channel $\Gamma_A\otimes\Gamma_B$ can be written in the form $p\ket{00}\bra{00}+(1-p)\ket{11}\bra{11}$ for which the entropy is upper bounded by $1$.

\section{Superactivation of $\mathcal{R}$ }

\begin{figure}
	\centering
		\includegraphics[scale=0.45]{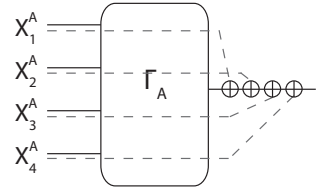}
	\caption{Construction of the channel $\tilde{\Gamma}$. The solid lines represent the qubits lines and the dashed represent the bit lines. The bit lines controls $\mathrm{CNOT}$s performed on the output of the channel. }
	\label{fig:tildegamma}
\end{figure}

We now turn to on the superactivation of the classical capacity regions of the channels $\tilde{\Gamma}_A,\tilde{\Gamma}_B$.
Namely we will show that if the senders can transmit only product states, classical capacity region $\mathcal{R}_{prod}=\mathcal{R}^{(1)} (\tilde{\Gamma}_A^{(\delta)})+\mathcal{R}^{(1)}(\tilde{\Gamma}_B^{(\delta)})$ is bounded by the inequalities $R_S=\sum_{i\in S}R_i \leq 2\delta$ for any subset of senders $S$.
We compare this with a protocol using entangled states. $\mathcal{R}_{ent}=\mathcal{R}^{(1)} (\tilde{\Gamma}_A^{(\delta)}\otimes\tilde{\Gamma}_B^{(\delta)})$ achieved in this case is given by inequality $R_1+R_2+R_3+R_4\leq 1$.

Let us present the channel $\tilde{\Gamma}_A$ (see FIG. \ref{fig:tildegamma}). It is a 4-to-1 channel. Input of each sender consists $d$-qubits line and $1$-bit line. The channel acts as:
\begin{eqnarray}
&&\tilde{\Gamma}_A^{(\delta)}(\rho_1\otimes\e^{(i)}_1\otimes\ldots\otimes\rho_4\otimes\e^{(l)}_4)\\
&=&\mathrm{CNOT}_{i}\circ\ldots\circ\mathrm{CNOT}_{l}( \Gamma_A^{(\delta)}(\rho_1\otimes\ldots\otimes\rho_4))\nonumber
\end{eqnarray}
where $\rho$ is transmitted through qubit inputs and $e^{(.)}$ through bit inputs. $\mathrm{CNOT}_0(\rho) = \rho$ and $\mathrm{CNOT}_1(\rho)=X\rho X^\dagger$. In the same way we construct $\tilde{\Gamma}_B^{(\delta)}$.

Note that $R_S\leq I(X_{S}:Y|X_{S^C})\leq H_{max}-H_{min}$ where $H_{max}$ is the maximal entropy of an output of a channel.
By the dimensionality of the output of the channels $\tilde{\Gamma}_A^{(\delta)}$ and $\tilde{\Gamma}_B^{(\delta)}$, we have in both cases $H_{max}\leq 1$. Taking into account results from the previous section we have that $H_{min} \geq 1-\delta$ that leads to $R_S(\tilde{\Gamma}_A^{(\delta)})\leq \delta$, $R_S(\tilde{\Gamma}_B^{(\delta)})\leq \delta$ and $R_S(\tilde{\Gamma}_A^{(\delta)})+R_S(\tilde{\Gamma}_B^{(\delta)})\leq 2 \delta$.

Now consider use of entangled states for communication. In this protocol each sender transmits the state $\Psi_{0,0}$ through the quantum lines, the label $0$ through the classical lines of the channel $\tilde{\Gamma}_A$ and with equal probability labels $0$ or $1$ through the classical lines of the channel $\tilde{\Gamma}_B$ . As noted above, outputs of the channels $\Gamma_A^{(\delta)}$ and $\Gamma_A^{(\delta)}$ are identical.
Perform the CNOT operation controlled by the output of $\tilde{\Gamma}_A$ on the output of $\tilde{\Gamma}_B$.
The result of CNOT pertaining to the channel $\tilde{\Gamma}_B$ and the classical input lines of this channel can be viewed as the output and input of the well known in the classical information theory binary XOR channel.
Its capacity region is given by the $R_1+R_2+R_3+R_4\leq 1$ \cite{xorchannel}.

\section{Conclusions}

We have shown that very strong subadditivity of the minimum output entropy $H_{min}$ and superadditivity of the capacity region $\mathcal{R}^{(1)}$ occurs in the domain of entanglement breaking quantum multiple access channels. The effect is bases on the fundamental properties of MAC i.e. independence of the senders.

Randomness extractors which extract from one randomness source requires additional random seed. That makes them useless for our purposes. The requirement of at least two randomness sources determines the number of senders in the example we presented.

We have shown that superadditivity effect for $R_T$ occurs for "single shot" capacity regions $\mathcal{R}^{(1)}$ of two different channels.
As it was shown in \cite{disc_supp} the superadditivity of the regularized classical capacity regions $\mathcal{R}^{(\infty)}$ of two different MACs occurs in realm of single user rates $R_i$, however the superadditivity of the regularized classical capacity $\mathcal{C}^{(\infty)}$ of 1-to-1 channels and superadditivity of $R_T$ of $\mathcal{R}^{(\infty)}$ of MACs still remains an open questions.
\section{Acknowledgments}

The author thanks P. Horodecki and R.W. Chhajlany for discussions and valuable comments on manuscript. This work was supported by Ministry of Since and Higher Education Grant No. NN202231937 and by the QESSENCE project.

\end{document}